# Cloud-based DDoS Attacks and Defenses

Marwan Darwish, Abdelkader Ouda, Luiz Fernando Capretz
Department of Electrical and Computer Engineering
University of Western Ontario
London, Canada
{mdarwis3, aouda, lcapretz}@uwo.ca

*Abstract* — Safety and reliability are important in the cloud computing environment. This is especially true today as distributed denial-of-service (DDoS) attacks constitute one of the largest threats faced by Internet users and cloud computing services. DDoS attacks target the resources of these services, lowering their ability to provide optimum usage of the network infrastructure. Due to the nature of cloud computing, the methodologies for preventing or stopping DDoS attacks are quite different compared to those used in traditional networks. In this paper, we investigate the effect of DDoS attacks on cloud resources and recommend practical defense mechanisms against different types of DDoS attacks in the cloud environment.

## I. INTRODUCTION

Cloud computing is the utilization of hardware and software combined to provide services to end users over a network like the internet. It includes a set of virtual machines that simulate physical computers and provide services, such as operating systems and applications. However, configuring virtualization in a cloud computing environment is critical when deploying a cloud computing system. A cloud computing structure relies on three service layers: Infrastructure as a Service (IaaS), Platform as a Service (PaaS), and Software as a Service (SaaS) (Fig. 1). IaaS gives users access to physical resources, networks, bandwidth, and storage. PaaS builds on IaaS and gives end users access to the operating systems and platforms necessary to build and develop applications, such as databases. SaaS provides end users with access to software applications.

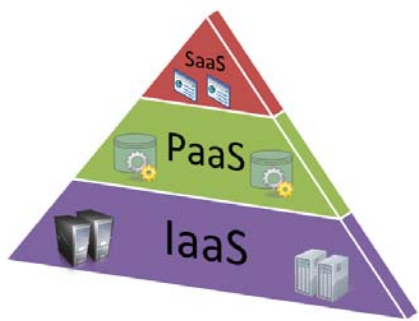

Figure 1. Cloud Computing Architecture

DDoS attacks are major security risks in a cloud computing environment, where resources are shared by many users. A DDoS attack targets resources or services in an attempt to render them unavailable by flooding system resources with heavy amounts of unreal traffic. The objective of DDoS attacks is to consume resources, such as memory, CPU processing space, or network bandwidth, in an attempt to make them unreachable to end users by blocking network communication or denying access to services. Dealing with DDoS attacks at all layers in cloud systems is a major challenge due to the difficulty of distinguishing the attacker's requests from legitimate user requests, even though the former come from a large number of distributed machines.

In this paper, we present an in-depth analysis of DDoS attacks in cloud computing and discuss the challenges in defending against these attacks. Section 2 provides an overview of DDoS attacks. Section 3 examines the effects of different types of DDoS attack and the recommended defense mechanisms for each type. Section 4 summarizes the results of investigations on DDoS attacks in the cloud system. Finally, Section 5 presents a brief summary of the paper.

## II. DDOS OVERVIEW

DDoS attacks have become more sophisticated. Many websites and large companies are targeted by these types of attacks. The first DDoS attack was reported in 1999 [1]. In 2000, large resource companies, such as Yahoo, Amazon, CNN.com and eBay, were targeted by DDoS attacks, and their services were stopped for hours [2]. Register.com was targeted by a DDoS in 2001; this was the first DDoS attack to use DNS servers as reflectors [3]. In 2002, service disruption was reported at 9 of 13 DNS root servers due to DNS backbone DDoS attacks. This attack recurred in 2007 and disrupted two DNS root servers. In 2003, Microsoft was targeted by Worm Blaster. One million computers were attacked by MyDoom in 2004. In 2007, a DDoS attack was carried out by thousands of computers, targeting more than 10,000 online game servers. In 2008, a DDoS attack targeting Wordpress.com caused 15 minutes of denial [4]. In 2009, GoGrid, a cloud computing provider, was targeted by a large DDoS attack, affecting approximately half of its thousands of customers. In 2009, Register.com was affected again by a DDoS attack. In the same year, some social networking sites, including Facebook and Twitter, were targeted by a DDoS. Many websites were attacked by DDoS in 2010, including the Australian Parliament House website, Optus, Web24, Vocus, and Burma's main Internet provider. In 2011, Visa, MasterCard, PayPal, and

PostFinance were targeted by a DDoS that aimed to support the WikiLeaks founder [4]. In the same year, the site of the National Election Commission of South Korea was targeted by DDoS attacks. Furthermore, thousands of infected computers participated in a DDoS attack that targeted the Asian E-Commerce Company in 2011 [4]. In 2012, the official website of the office of the vice-president of Russia was unavailable for 15 hours due to a DDoS attack [4]. In the same year, many South Korean and United States (US) websites were targeted by a DDoS. Godaddy.com websites reported service outages because of such an attack. In 2012, major US banks and financial institutions became the target of a DDoS attack.

DDoS attacks are evolving rapidly and are targeting large companies, which cause huge financial losses to those companies and websites globally. Consequently, investigating DDoS attacks in the cloud system is vital along with recommending mechanisms to mitigate such attacks.

## III. DDoS ATTACKS AND DEFENSES

DDoS attacks affect all layers of the cloud system (IaaS, PaaS, and SaaS) and can occur internally or externally. An external cloud-based DDoS attack starts from outside the cloud environment and targets cloud-based services. This type of attack affects the availability of services. The most affected layers in the cloud system by an external DDoS attack are the SaaS and PaaS layers. An internal cloud-based DDoS attack occurs within the cloud system, primarily in the PaaS and IaaS layers, and can occur in several ways. For example, the attackers may take advantage of the trial periods of cloud services of some vendors. As a result, an authorized user within the cloud environment can launch a DoS attack on the victim's machine internally. On the other hand, sharing infected virtual machine images could allow an attacker to control and use the infected virtual machines to carry out an internal DDoS attack on the targeted machine within the same cloud computing system. A DDoS includes different types of attacks. Descriptions of those attacks and recommended practical defense mechanisms in the cloud system are presented in the following sections.

### A. IP spoofing attack

In the Internet Protocol (IP) spoofing attack, packet transmissions between the end user and the cloud server can be intercepted and their headers modified such that the IP source field in the IP packet is forged by either a legitimate IP address, as shown in Fig. 2, or by an unreachable IP address. As a result, the server will respond to the legitimate user machine, which affects the legitimate user machine, or the server will be unable to complete the transaction to the unreachable IP address, which affects the server resources. Tracing such an attack is difficult due to the fake IP address of the IP source field in the IP packet. The methods for detecting an IP spoofing attack can be applied in the PaaS layer or in the network resources on the IaaS layer.

Due to the difficulty of modifying and upgrading different types of network resources in the cloud system, hop-count filtering (HCF) [5] can be used to distinguish legitimate IPs from spoofed IPs in the PaaS layer. The HCF counts the number of hops depending on the value of the Time to Live (TTL) field in the IP header. IP-to-hop-count (IP2HC) mapping is built to detect the spoofed packet. An analysis concluded by Wang et al. [5] indicated that 90% of spoofed addresses can be detected using the HCF method. One drawback of this method is that attackers can build their own IP2HC mapping to avoid HCF. A trust-based approach to detect spoofed IP addresses can be used in the access routers on the IaaS layer [6], but another compatible solution should be proposed to detect IP spoofing in distribution routers.

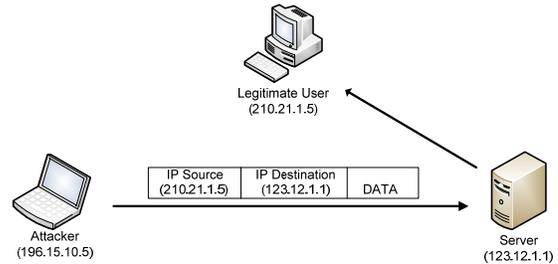

Figure 2. IP spoofing attack

### B. SYN flooding attack

A Transmission Control Protocol (TCP) connection starts with a three-way handshake, as shown in Fig. 3(a). A typical three-way handshake between a legitimate user and the server begins by sending a connection request from the legitimate user to the server in the form of a synchronization (SYN) message. Then, the server acknowledges the SYN by sending back (SYN-ACK) a request to the legitimate user. Finally, the legitimate user sends an ACK request to the server to establish the connection. SYN flooding occurs when the attacker sends a huge number of packets to the server but does not complete the process of the three-way handshake. As a result, the server waits to complete the process for all of those packets, which makes the server unable to process legitimate requests, as shown in Fig. 3(b). Also, SYN flooding can be carried out by sending packets with a spoofed IP address. A sniffing attack is considered a type of SYN flooding attack. In a sniffing attack, the attacker sends a packet with the predicted sequence number of an active TCP connection with a spoofed IP address. Thus, the server is unable to reply to that request, which affects the resource performance of the cloud system.

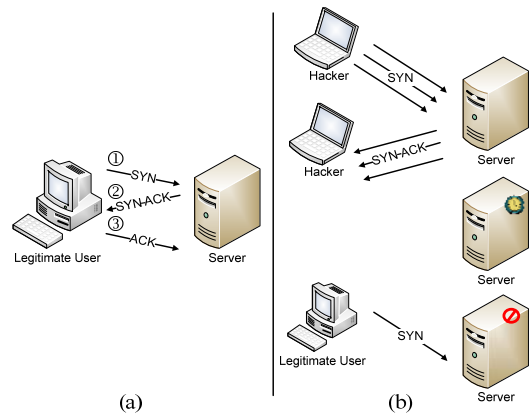

Figure 3. SYN flooding attack

Many defense mechanisms against SYN flooding attack can be used in the PaaS and IaaS layers [7]. The SYN cache approach [8], which establishes a connection with a legitimate request, can be considered in the PaaS layer, but this causes an increase in latency by 15%. The SYN cookies defense mechanism [8] is another recommended defense mechanism in the PaaS layer to detect a SYN flooding attack, but it lowers the performance of the cloud system. Reducing the time of the SYN received to degrade the timeout is a recommended PaaS defense measure, but legitimate ACK packets could be lost. Moreover, some detection mechanisms, including filtering, firewall, and active monitoring, can be used in the IaaS layer. Filtering is an effective method to prevent a SYN flooding attack by configuring internal and external router interfaces, but this method is not reliable due to its limited use. The firewall mechanism in the IaaS layer depends on splitting the TCP connection, but this could affect the performance of the networking system. An active monitoring mechanism [9] can be used in the IaaS layer to monitor traffic of the TCP/IP and react in cases of SYN flooding. However, this approach depends on the SYN cookies mechanism, which leads to decreased performance of cloud resources.

### C. Smurf attack

In a smurf attack, the attacker sends a large number of Internet Control Message Protocol (ICMP) echo requests. These requests are spoofed such that its source IP address is the victim's IP, and the IP destination address is the broadcast IP address, as shown in Fig. 4. As a result, the victim will be flooded with broadcasted addresses. The worst case occurs when the number of hosts who reply to the ICMP echo requests is too large. Preventing this type of attack is difficult, but it can be mitigated by two different mechanisms. The first recommended defense mechanism in the IaaS layer is configuring the routers to disable the IP-directed broadcast command; this is disabled by default in current routers. However, the attacker could use the compromised device in the cloud system as an intermediary to send ICMP echo requests to the broadcast IP address locally, thereby carrying out an internal cloud-based DoS attack. Configuring the router in the IaaS layer cannot prevent a smurf attack. Consequently, a second defense mechanism is needed, which is configuring the operating systems in the PaaS layer so that there is no response to the ICMP packets sent to the IP broadcast addresses.

### D. Buffer overflow attack

In a buffer overflow attack, the attacker sends an executable code to the victim in order to take advantage of buffer overflow vulnerability. As a result, the victim's machine will be controlled by the attacker. The attacker could either harm the victim's machine or use the infected machine to perform an internal cloud-based DDoS attack. Four defense mechanisms to prevent buffer overflow vulnerability can be used in the SaaS layer [10].

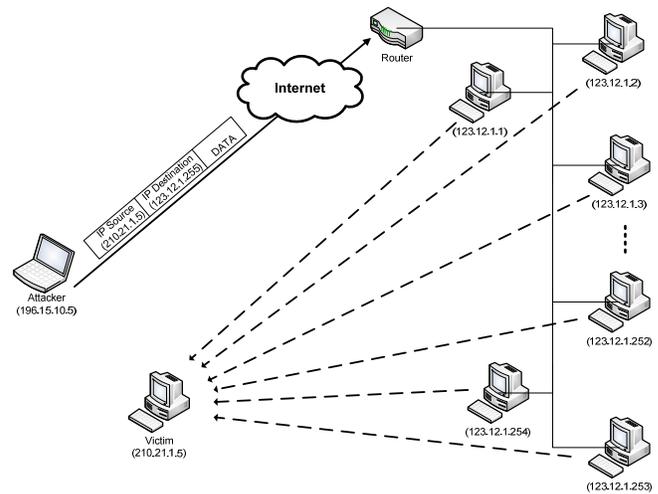

Figure 4.  Smurf attack

The first mechanism is preventing such vulnerability when writing the source code [11]; however, time consumption is a limitation. Performing a check of the array bounds is a second recommended defense mechanism; this consists of checking the memory access and compiler and using safety language. The third defense mechanism is runtime instrumentation, which can either modify the return address to detect the vulnerability or estimate the buffer bounds then perform a check of the runtime bounds. The fourth recommended defense mechanism in the SaaS layer is analyzing the static and dynamic code to detect application vulnerability in this layer.

### E. Ping of death attack

In the ping of death attack, the attacker sends an IP packet with a size larger than the limit of the IP protocol, which is 65,535 bytes, as shown in Fig. 5. Handling an oversized packet affects the victim's machine within the cloud system as well as the resources of the cloud system. Recent network resources and operating systems disregard any IP packets larger than 65,535 bytes. Therefore, such attacks are not currently affecting any cloud system layers.

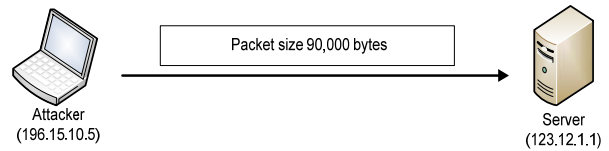

Figure 5.  Ping of death attack

### F. Land attack

This attack uses the "Land.c" program to send forged TCP SYN packets with the victim's IP address in the source and destination fields, as shown in Fig. 6. In this case, the machine will receive the request from itself and crash the system. Such an attack is prevented in recent networking devices and operating systems by dropping ICMP packets that contain the same IP address in the source and destination fields. Consequently, there is no need for a land attack defense mechanism to be used in all layers of the cloud system.

However, the process of checking and dropping large amounts of ICMP requests could affect the resources of the victim's machine in the PaaS layer or the networking resources in the IaaS layer.

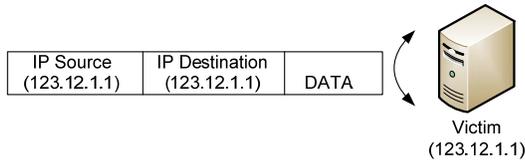

Figure 6. Land attack

## G. Teardrop attack

This kind of attack uses the "Teardrop.c" program to send invalid overlapping values of IP fragments in the header of TCP packets. As a result, the victim's machine within the cloud system will crash in the re-assembly process. Recent operating systems and network resources can handle such attacks. Therefore, teardrop attacks no longer affect any layers of cloud computing.

## IV. SUMMARY OF CLOUD-BASED DDoS ATTACK

Based on an investigation of the major types of DDoS attacks, we derive a taxonomy of cloud-based DDoS attacks, as illustrated in Table 1. Several taxonomies of DDoS attacks exist [4] [12] [13]. Our classification is focused on cloud computing aspects, such as a cloud-based type of attack, recommended practical defense mechanisms, and the drawbacks of each defense mechanism.

TABLE I. TYPES OF DDoS ATTACKS ON THE CLOUD SYSTEM

| Attack | Cloud-based type | Recommended Practical Defense Mechanism | Drawback |
|---|---|---|---|
| IP spoofing | External Internal | - Hop Count Filtering (HCF) in the PaaS layer [5] | - The attacker can build his own IP2HC mapping to avoid HCF |
| | | - Trust-based approach in the IaaS layer [6] | - Another compatible solution should be proposed to detect IP spoofing in distribution routers |
| SYN flooding | External Internal | - SYN cache approach in the PaaS layer [8] | - Increase in latency |
| | | - SYN cookies defense approach in the PaaS layer [8] | - Lowers the performance of the cloud system |
| | | - Reducing the time of SYN-received in the PaaS layer | - Some of the legitimate ACK packets could be lost |
| | | - Filtering mechanism in the IaaS layer | - Not reliable due to the limited use of this method |
| | | - Firewall mechanism in the IaaS layer | - May affect the performance of the networking system in the cloud |
| | | - Active monitoring mechanism in the IaaS layer [9] | - Decreases resource performance in the cloud |
| Smurf attack | External Internal | - Configuring virtual machines in the PaaS layer | |
| | | - Configuring network resources in the IaaS layer | |
| Buffer overflow | External Internal | - Preventing when writing source code mechanism in the SaaS layer [10] | - Time consumption |
| | | - Performing the array bounds checking mechanism in the SaaS layer [10] | |
| | | - Runtime instrumentation mechanism in the SaaS layer [10] | |
| | | - Analyzing the static and dynamic code mechanism in the SaaS layer [10] | |
| Ping of death | External Internal | - Difficult to affect any layers of the cloud system currently, but the attack could be developed in the future | |
| Land.c | External Internal | - Difficult to affect any layers of the cloud system currently, but the attack could be developed in the future | |
| Teardrop.c | External Internal | - Difficult to affect any layers of the cloud system currently, but the attack could be developed in the future | |

## V. CONCLUSION

DDoS attacks are currently a major threat and work against the availability of cloud services. With each developed defense mechanism against DDoS attacks, an improved attack appears. Defense mechanisms to protect against DDoS attacks are not always effective on their own. Combining different mechanisms to build hybrid defense mechanisms, in particular with different cloud computing layers, is highly recommended. It is extremely important to investigate the effects of these different types of DDoS attacks on the cloud system. In this paper, historical examples of DDoS attacks have been presented. We also investigated the effect of different types of DDoS attacks on the cloud environment. Finally, we analyzed and identified recommended defense mechanisms against DDoS attacks in cloud-based systems.

ACKNOWLEDGMENT

This work was partially supported by King Abdulaziz University through the Cultural Bureau of Saudi Arabia in Canada. This support is greatly appreciated.